\documentclass[preprint,
 amsmath,amssymb,
 aps,
]{revtex4-2}

\usepackage[utf8]{inputenc}
\usepackage{graphicx}
\usepackage{amsmath}
\usepackage{siunitx}
\usepackage{xcolor}

\begin{document}

\title{Mixing in confined fountains} 

\author{You-An~Lee}
\affiliation{Physics of Fluids Group and Max Planck
Center Twente for Complex Fluid Dynamics, University of
Twente, Enschede, The Netherlands}

\author{Detlef~Lohse}
\affiliation{Physics of Fluids Group and Max Planck
Center Twente for Complex Fluid Dynamics, University of
Twente, Enschede, The Netherlands}
\affiliation{Max Planck Institute for Dynamics and Self-Organization, Am Fa{\ss}berg 17, G\"ottingen, Germany}

\author{Sander~G.~Huisman}
\affiliation{Physics of Fluids Group and Max Planck
Center Twente for Complex Fluid Dynamics, University of
Twente, Enschede, The Netherlands}


\date{\today}

\begin{abstract}
We have experimentally investigated mixing in highly confined turbulent fountains, namely quasi-two-dimensional fountains. Fountains are formed when the momentum of the jet fluid is in the opposite direction to its buoyancy force. This work consists of two parts. First, we injected an ethanol/oil mixture (ouzo mixture) downward into quiescent water, forming a quasi-2D fountain with oil droplet nucleation (ouzo fountain). In the steady state, nucleation is restricted to the fountain rim, and there is hardly any nucleation in the fountain body, suggesting limited mixing with the bath in the quasi-two-dimensional fountain. By injecting a dyed ethanol solution as a reference case, we confirmed that the local water fraction within the fountain is indeed insufficient to induce nucleation. 

Second, we have studied the effect of density difference between the jet fluid and the ambient water systematically. We injected saline solutions upward into quiescent water with various concentrations of sodium chloride (NaCl) at various flow rates. The fountains show stronger mixing and thus lower concentration in the initial negatively buoyant jet (NBJ) stage. In the steady fountain stage, the confinement induces the shielding effect by the outer flow, which reduces the degree of mixing and leads to higher concentrations. Also, we show that the density difference is the critical parameter that determines the fountain concentration. The decreasing concentration with the density difference indicates that the larger (negative) buoyancy effect enhances the stretching of the fluid parcels \citep{Villermaux2019}, leading to a higher degree of mixing in the fountain. From the probability density functions of the concentration, we demonstrate that the degree of mixing in the steady fountain stage is largely determined in the developing stages for a quasi-2D fountain.   
\end{abstract}
{\let\clearpage\relax\maketitle}
\section{Introduction}
Turbulent fountains have been extensively studied for their applications in industry and natural occurrences \citep{Hunt2015}. Fountains are formed when the momentum of the jet fluid is in the opposite direction to its buoyancy force. The flow first goes through an initial transient stage, that is, the negatively buoyant jet (NBJ) stage. In the NBJ stage, the flow behaves like a positively buoyant jet (PBJ), entraining ambient fluid and expanding laterally. When the buoyancy force overcomes the momentum, the flow starts to reverse from its initial peak height, forming a coaxial structure consisting of an inner NBJ region and an outer PBJ region. After a developing period, the flow reaches a (quasi-)steady fountain stage, with the fountain top fluctuating around a steady state height. A major focus of the previous studies on turbulent fountains is the scaling laws of the shape parameters and the source Froude number $Fr_0=u_0/(\sqrt{rg\Delta \rho/\rho_0})$, where $u_0$ is the source velocity, $r$ the needle radius, $\nu$ the kinematic viscosity, $g$ the gravity, $\Delta \rho=\rho_{\text{jet}}-\rho_0$ the density difference between the jet fluid and the ambient water, and $\rho_0$ the density of the ambient water. The shape parameters describing the resulting fountain include the mean rise heights \citep{Williamson2011, Burridge2012, Hunt2015}, the rise height ratio \citep{Williamson2011, Burridge2012, Hunt2015}, and the fluctuating frequency of the fountain top \citep{Burridge2013}. A turbulent fountain can be categorized using scaling laws for the shape parameters \citep{Hunt2015}.   

In addition to the shape parameters, entrainment and mixing of a turbulent fountain are crucial but receive less attention due to the complex structure of a fountain. In recent years, more and more efforts have been devoted to this aspect. \citet{Burridge2016} estimated entrainment in a fountain from a global perspective, showing the entrained volume flux as a function of $Fr_0$. \citet{Milton2022} characterized the structure of a fountain using its mean velocity and concentration profiles and the relevant turbulent statistics. They \citep{Milton2022} established a framework to analyze the entrainment between the inner NBJ flow and the outer reverse flow. \citet{Talluru2022} investigated entrainment and dilution in the fountain cap using simultaneous velocity and concentration measurements. They showed that the entrainment does not equal the scalar dilution in the fountain cap, and the local Reynolds number at the base of the cap governs dilution. \citet{Xue2019} conducted fountain experiments in a filling box setup \citep{Baines1969}, where the reversed flow accumulates from the bottom (or the top). They quantified the degree of entrainment and mixing using the thickness of the accumulated dyed fluid from the reversed flow, which is shown to be a function of source conditions $Re_0$ and $Fr_0$. 

While the aforementioned studies addressed 3D round fountains, there have also been efforts to explore fountains with different source geometry \citep{Hunt2019} and under confinement \citep{Debugne2018}. \citet{Hunt2019} studied line fountains released from a high aspect ratio rectangular slot. Tracking the shape evolution, they \citep{Hunt2019} obtained the scaling law for the rise height and the source Froude number, and classified the flow based on its lateral flapping behavior. \citet{Debugne2018} looked into fountains with different levels of spanwise confinement. They \citep{Debugne2018} proposed a regime map based on the visualization to categorize the flow. The fountains are categorized based on the confinement ratio, $W/r_0$, and the confined Froude number, $Fr_c = Fr_0(W/r_0)^{-1.25}$, where $W$ is the gap width of the confinement, and $r_0$ the radius of the needle. These works \citep{Debugne2018,Hunt2019}, however, did not address the effect of source geometry or confinement on the degree of mixing. 

Mixing is an essential topic for fluid dynamics research \cite{Rothstein1999, Duplat2008, Villermaux2019}, which determines the structure of the flow field and the temporal evolution of concentration distribution. Using a lamellar representation, Villermaux pointed out that mixing is in fact an enhanced diffusion process by stirring or stretching \cite{Villermaux2019}. There are various ways of stretching, which reduces the scale of the lamellar down to the Batchelor scale \cite{Batchelor1959}, where molecular diffusion sets in and dominates the mixing process. For a multicomponent fluid, mixing also has a profound effect on the accompanying processes, such as chemical reaction \cite{Mingotti2019b, Guilbert2021a, Guilbert2021b} or solvent exchange \cite{Lee2022}.

Solvent exchange in Hele-Shaw-like microfluidic setups has been extensively studied in the laminar regime \cite{Zhang2015,Hajian2015, Lu2015, Lu2016,Lu2017, Li2018,Dyett2018,Zeng2019,Li2021}. These works focused on the heterogeneous nucleated surface nanodroplets, calculating their total mass, their droplet size distribution, and the scaling laws relating the amount of nucleation to the flow parameters. However, solvent exchange in the turbulent regime remains almost unexplored.

In this paper, we aim to quantify mixing in a highly-confined fountain \citep{Debugne2018}. We first introduce the experimental setup, the methods to measure the concentration, and the experimental conditions. Then we present the results of solvent exchange in a highly-confined fountain. As the density difference is fixed for the ethanol-dominated ouzo mixture, we further conduct experiments using saline solutions of various compositions, revealing the effect of density difference on mixing in the second part of the study.

\section{Experiment}
\subsection{Set-up}

\begin{figure}[h!tb]
\centering
\includegraphics[scale=1]{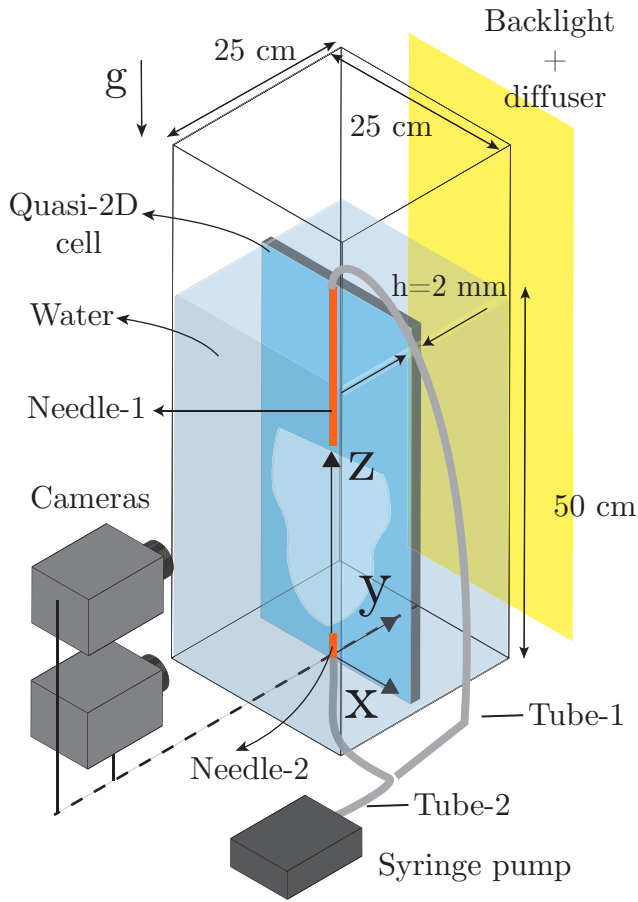}
\vspace{-5mm}
\caption{Experimental set-up. For the downward injection of dyed ethanol and ouzo mixture, we used Tube-1 and Needle-1. For the upward injection of dyed saline solutions, we switched to Tube-2 and Needle-2.}
\label{img:setup4}
\end{figure}

The experimental setup in Fig.\ \ref{img:setup4} is similar to what we used in Chapters 2 and 3, namely a quasi-2D geometry with a \SI{2}{\mm} gap immersed in a big water tank with dimensions \SI{25}{\cm} $\times$ \SI{25}{\cm} $\times$ \SI{50}{\cm} (W $\times$ L $\times$ H). In the first part of our study, we injected an ouzo mixture downwards into the quiescent water in the tank through a round needle with inner diameter  \SI{0.51}{\mm}, outer diameter \SI{0.82}{\mm}, and length \SI{300}{\mm} from the top. The ouzo mixture consists of ethanol and trans-anethole (Sigma Aldrich, $\geq$99\%) with a weight ratio $w_e:w_o=100:1$. As a reference experiment, we also injected dyed ethanol, which has an almost identical density to the ouzo mixture. We used red food dye from JO-LA. The fluids were injected through a Harvard 2000 syringe pump at \SI{100}{\ml/\min}. In the second part of our study, we injected saline solutions upwards into the tank through a round needle with inner diameter \SI{0.51}{\mm}, outer diameter \SI{0.82}{\mm}, and length \SI{12.7}{\mm} at various flow rates to reach different regimes spanned by the source Reynolds number $Re_0=u_0 r/\nu$ and the source Froude number $Fr_0=u_0/(\sqrt{rg\Delta \rho/\rho_0})$, where $u_0$ is the source velocity, $r$ the needle radius, $\nu$ the kinematic viscosity, $g$ the gravity, $\Delta \rho=\rho_{\text{jet}}-\rho_0$ the density difference between the jet fluid and the ambient water, and $\rho_0$ the density of the ambient water. The experimental conditions and the resulting $Re_0$ and $Fr_0$ are listed in Table \ref{tbl:condition4}.

\begin{table}[h!bt]
\begin{center}
\def\arraystretch{1.3}
\setlength{\tabcolsep}{7pt}
\begin{tabular}{r r r r r r r} 
Exp.\ & Fluid & $\Delta \rho/\rho$ & \vtop{\hbox{\strut $Q$ $(\text{m}^{\text{3}}/\text{s})$}\hbox{\strut $\times 10^{-7}$}} & $Re_0$ & $Fr_0$ & $n$\\
A1 & ouzo & -0.22 & 16.67 & 472 & 23 & 2\\ 
A2 & dyed ethanol & -0.22 & 16.67 & 472 & 23 & 2\\ 
B1 & saline & 0.01 & 5.00 & 227 & 33 & 2\\ 
B2 & saline & 0.01 & 8.33 & 378 & 55 & 2\\
B3 & saline & 0.03 & 5.00 & 228 & 19 & 2\\
B4 & saline & 0.03 & 8.33 & 380 & 32 & 2\\
B5 & saline & 0.03 & 16.67 & 760 & 63 & 2\\
B6 & saline & 0.05 & 5.00 & 228 & 15 & 3\\ 
B7 & saline & 0.05 & 8.33 & 379 & 25 & 3 \\ 
B8 & saline & 0.05 & 16.67 & 758 & 49 & 4\\
B9 & saline & 0.09 & 5.00 & 222 & 11 & 6\\
B10 & saline & 0.09 & 8.33 & 370 & 18 & 3\\
B11 & saline & 0.09 & 16.67 & 740 & 37 & 4\\
B12 & saline & 0.10 & 8.33 & 366 & 17 & 5 \\ 
B13 & saline & 0.10 & 16.67 & 733 & 35 & 4 \\ 
B14 & saline & 0.15 & 8.33 & 339 & 14 & 6 \\
B15 & saline & 0.15 & 16.67 & 678 & 28 & 6\\
B16 & saline & 0.20 & 8.33 & 305 & 12 & 3 \\
B17 & saline & 0.20 & 16.67 & 609 & 24 & 5 \\
\end{tabular}
\end{center}
\caption{Experimental conditions. Exp.\ A1 and A2 are the focus of the first part of our study, while Exp.\ B1--B17 detail the conditions of the saline experiments in the second part of the study. $n$ in the last column shows the repetitions of experiments.}
\label{tbl:condition4}
\end{table}

We visualize the flow and measure the concentration field with a light attenuation technique, which requires a backlit optical setting as shown in Fig.\ \ref{img:setup4}. The light attenuation technique relies on an in-situ calibration to convert the recorded light intensity field to the concentration field. See Appendix A for more details about the calibration. We recorded the flow field using two Photron FASTCAM Mini AX200 high-speed cameras with Zeiss \SI{100}{\mm} objectives. The images were recorded with a 1024 $\times$ 1024 pixels resolution at 50 fps. The experiment for each condition is repeated 2--6 times for reliable statistical results. Although we took great care aligning the injection needle, the fountains with larger density differences ($\Delta \rho/\rho \geq 0.10$) tend to fall to one side when there is a small misalignment. We stopped the experiments as long as the fountain fell asymmetrically toward one side, and exclude those data from the statistical analysis.
\subsection{Oversaturation}

\begin{figure}[h!tb]
\centering
\includegraphics[scale=1.0]{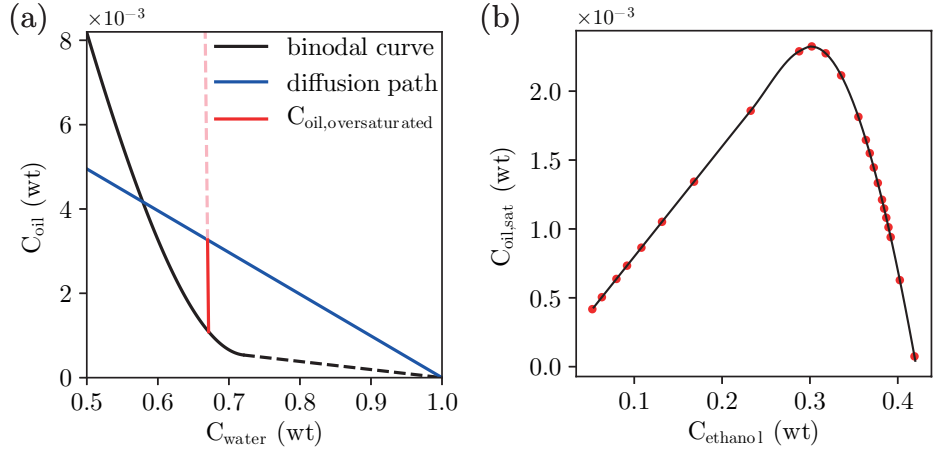}
\caption{Estimation of oversaturation. (a) Binodal curve and the theoretical diffusion path for ouzo mixture ${w_e/w_o = 100}$. Note that the binodal curve is determined partly by titration experiments (black solid line) and partly by direct linear approximation (black dashed line). The length of the red line segment measures the oversaturation. The pink dashed line extends to the red segment to the pure oil phase, which can be approximated with a vertical line considering the tiny amount of oil. (b) Non-monotonic variation of oversaturation as a function of ethanol weight fraction, which is fitted by a $3^{rd}$-order polynomials for $C_{\text{ethanol}}>0.3$, and by a piecewise linear function for $C_{\text{ethanol}}\leq0.3$.}
\label{img:phasegram}
\end{figure}

The concentration of the nucleated oil induced by the ouzo effect can be estimated using oversaturation as a function of the local water fraction, see Fig.\ \ref{img:phasegram}. The binodal curve in Fig.\ \ref{img:phasegram}(a) consists of two parts. The solid black curve is determined by titration, marking the saturation points for various compositions of the ternary liquid system. The black dashed line, on the other hand, is a linear approximation. We use such an approximation due to the difficulty in titration with low oil fraction. The diffusion path denotes the composition of the local fluid parcel upon mixing, which is pre-determined by the initial ethanol/oil ratio, and is assumed to be a straight line \citep{Li2021,Lee2022}. The difference between the diffusion path and the bimodal curve measures the oversaturation of the oil, which is extracted and displayed in Fig.\ \ref{img:phasegram}(b) as a function of ethanol fraction. Note that with $w_e:w_o=100:1$, the summation of the water fraction and the ethanol fraction can be considered as 1. Fig.\ \ref{img:phasegram}(b) will be used later with Fig.\ \ref{img:coreeth}(c) to predict the occurrence of the ouzo effect.  

The approach we use here is the same as our previous work for turbulent jet \citep{Lee2022}, which is based the previous efforts of solvent exchange in the laminar regime \citep{Zhang2015,Lu2017,Li2021}. However, we would like to point out that the oil droplets in turbulent flows nucleate mainly in the bulk due to intense turbulent mixing, while for the laminar cases, diffusion-triggered nucleation mainly occurs on the wall and at the front. Also, while the laminar studies calculated the total mass of the nucleated oil, we focus on (the temporal evolution of) the oil concentration.

\section{Analysis and Results}
\subsection{Ouzo and dyed ethanol fountains}

\begin{figure}[h!tb]
\centering
\includegraphics[scale=1]{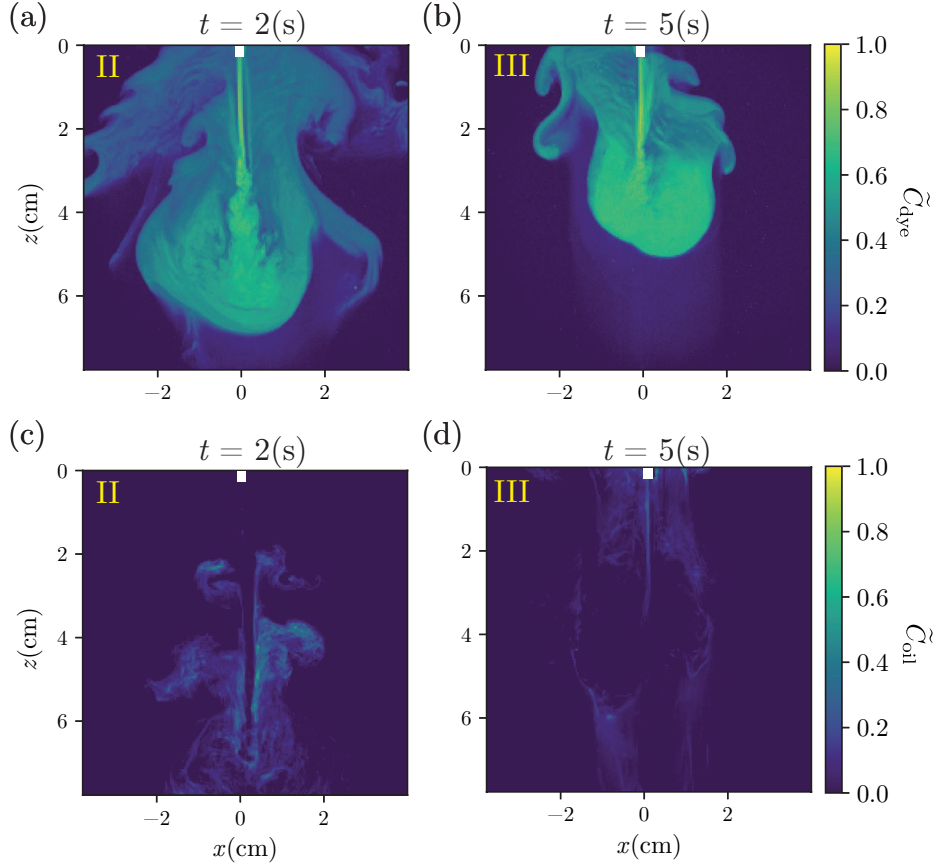}
\caption{Visualization of (a,b) the dyed fountain and (c,d) the ouzo fountain. (a,c) show the snapshots at $t=2$(s), which lies in the flow reversal regime (II). (b,d) display the snapshots at $t=5$(s) in the steady fountain regime (III). The needle top is marked in white.}
\label{img:cinst}
\end{figure}

As in a round fountain, a quasi-2D fountain also goes through three stages, (I) the NBJ stage, (II) the flow reversal, and (III) the (quasi-)steady fountain stage. Due to the very long needle used in the experiment, the transition to turbulence was delayed, the flow is fully laminarized, delaying the transition to turbulence in the NBJ stage. In Fig.\ \ref{img:cinst}(a,c) we show the flow reversal stage for the dyed ethanol (a) and the ouzo mixture (c). Mixing in this stage leads to an obvious dilution of the dyed fluid in Fig.\ \ref{img:cinst}(a), which corresponds to the nucleation of oil in Fig.\ \ref{img:cinst}(c). In the steady fountain stage, Fig.\ \ref{img:cinst}(d) shows that there is hardly any nucleation in the fountain core. This surprising finding suggests that the entrained water is restricted to the outer rim of the fountain, and does not mix sufficiently with the ethanol/oil mixture in the jet fluid. The corresponding reference dye case in Fig.\ \ref{img:cinst}(b) also shows a relatively high dye (ethanol) concentration, indicating weak mixing and dilution.

The normalized concentrations displayed in the color code in Fig.\ \ref{img:cinst} are defined as:

\begin{align}
& \widetilde{C}_{\text{dye}} = \frac{C_{\text{dye}}}{C_0}, \label{eq:cnormdye}\\
& \widetilde{C}_{\text{oil}} = \frac{C_{\text{oil,oversat}}-C_{\text{thresh}}}{C_{\text{max}}-C_{\text{thres}}}, \label{eq:cnormoil}
\end{align}
where $\widetilde{C}_{\text{dye}}$ can be considered as the ethanol fraction, $C_0$ the initial dye concentration \SI{3500}ppm, $C_{\text{oil,sat}}$ the oil oversaturation, $C_{\text{max}}$ the theoretical maximum oversaturation shown in Fig.\ \ref{img:phasegram}(b), and $C_{\text{thres}}$ the threshold oversaturation triggering nucleation, whose definition is detailed in Appendix A.

\begin{figure}[h!tb]
\centering
\includegraphics[scale=1]{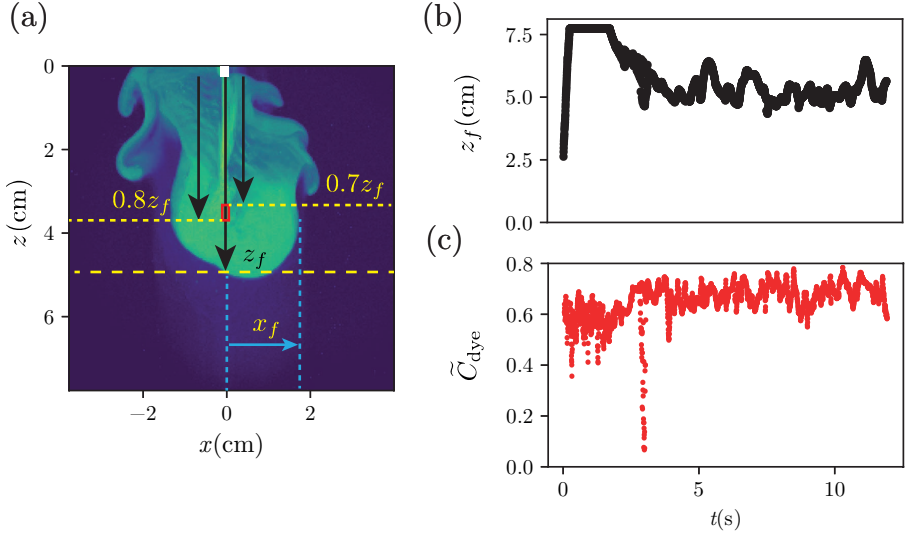}
\caption{(a) marks the definition of the parameters used in the analysis. The red rectangle encloses the defined fountain core with a width of 5 pixels, which is the image unit when converting the recorded light intensity to concentration. The needle top is marked in white. (b) tracks the temporal variation of the rising height of the fountain. The fountain top travels out of sight in the NBJ stage, leading to the flattened value at around $z_f=$\SI{7.5}{cm}. (c) shows the temporal evolution of the core concentration for the dyed ethanol fountain. Note that the sudden drop of $\widetilde{C}_{\text{dye}}$ at around $t=3$(s) was caused by a large unbalanced flapping motion toward one side of the fountain during flow reversal, which quickly resumed the quasi-steady state.}
\label{img:coreeth}
\end{figure}

Tracking the temporal evolution of the ethanol concentration in Fig.\ \ref{img:cinst}(b), we define the fountain core as a block positioned at 0.7$z_f$--0.8$z_f$ at the centerline, where $z_f$ is the position of the fountain top at the centerline, see Fig.\ \ref{img:coreeth}(a,b). The choice of 0.7$z_f$--0.8$z_f$ is to avoid the laminar jet at the bottom and on the other hand the fluctuating fountain top. Fig.\ \ref{img:coreeth}(c) shows the temporal variation of the normalized dye concentration (ethanol local fraction) of the core in a quasi-2D fountain in Fig.\ \ref{img:cinst}(a,b), which demonstrates that the ethanol local fraction is above 0.6 in the steady fountain stage. From Fig.\ \ref{img:phasegram}(b), we can easily confirm that nucleation can not be triggered for ethanol local fraction above 0.42, which is indeed the case displayed in Fig.\ \ref{img:cinst}(d).

\begin{figure}[h!tb]
\centering
\includegraphics[scale=1]{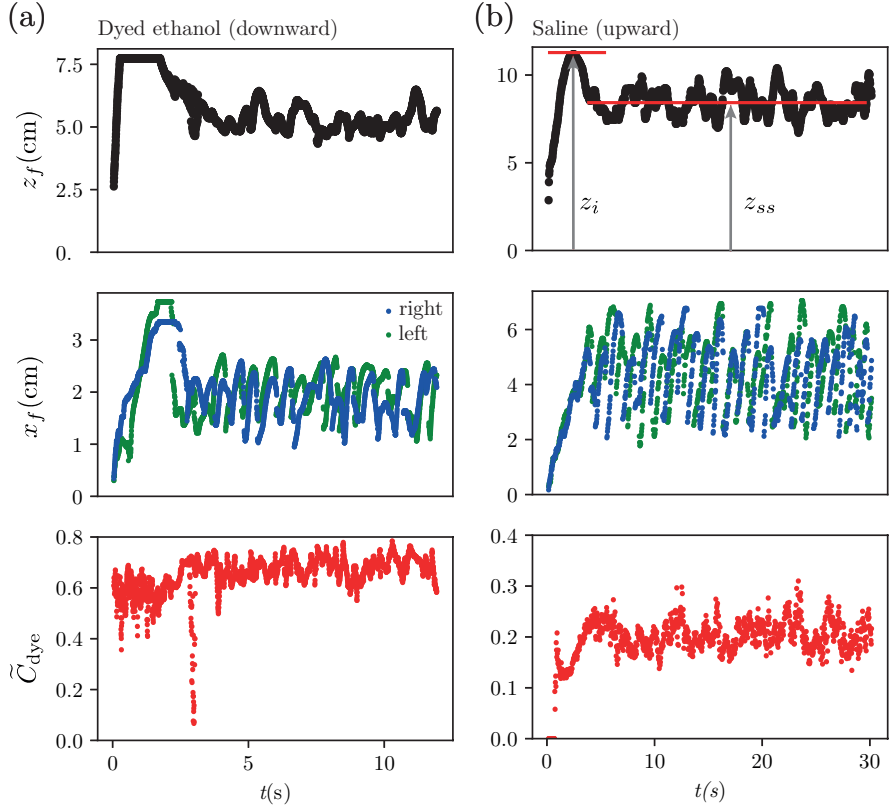}
\caption{Temporal evolution of the fountain height, width, and the core concentration for (a) the dyed ethanol fountain and (b) the saline fountain with density difference similar to the dyed ethanol one. (a) is the same as Figs.\ \ref{img:coreeth}(b,c). Note that the fountain is downward in (a) and upward in (b), and the coordinates are flipped based on the injection points.}
\label{img:updown}
\end{figure}

In the first part of the experiment, we show that mixing is highly limited in a quasi-2D fountain with ethanol as the dominant component. The ethanol-dominated solution in water gives a fixed density difference of around \SI{-22}{\%}. To obtain the effect of the density difference on mixing, we conduct experiments using saline solutions of various compositions. Before diving into this second part of the experiments, we compare the results of the ethanol fountain and the saline fountain with a similar density difference, $Re_0$, and $Fr_0$, that is, Exps.\ A2 and B17 in Table \ref{tbl:condition4}. 

In addition to a small mismatch of $Re_0$, the major differences between Exps.\ A2 and B17 are the direction of injection and the length of the injection needle. Fig.\ \ref{img:updown} presents the difference between these two sets of experiments in the fountain top position $z_f$, the lateral excursion $x_f$, and the core concentration $\widetilde{C}_{\text{dye}}$. The definition of $z_f$ and $x_f$, and the fountain core concentration can be found in Fig.\ \ref{img:coreeth}(a). Fig.\ \ref{img:updown} exhibits significant differences between Exps.\ A2 and B17 in all three parameters. Although \citet{Vaux2019} already discussed the difference between the upward and downward non-Boussinesq turbulent fountains, we can't attribute the direction of injection as the case for this significant difference as was proposed in \citet{Vaux2019}. 

As mentioned earlier, for the ouzo and the dyed ethanol, we used a long injection needle, which leads to a delayed transition to turbulence and a laminar NBJ stage. The reason why we use the long needle is to prevent the accumulated reverse flow from interfering with the fountain body. The delayed transition to turbulence leads to significantly reduced $z_f$, $x_f$, and the degree of mixing in the fountain core. The low concentration in the last panel of Fig.\ \ref{img:updown}(b) suggests stronger mixing if the jet becomes turbulent without delay. Therefore, we emphasize that the pronounced differences between Figs.\ \ref{img:updown}(a) and \ref{img:updown}(b) actually result from the injection geometry. The poor mixing of the laminar NBJ stage for the ouzo and the dyed ethanol fountains causes directly restricted mixing in the fountain stage in Figs.\ \ref{img:cinst}(b,d). 
\subsection{Saline fountains}

\begin{figure}[h!tb]
\centering
\includegraphics[scale=1]{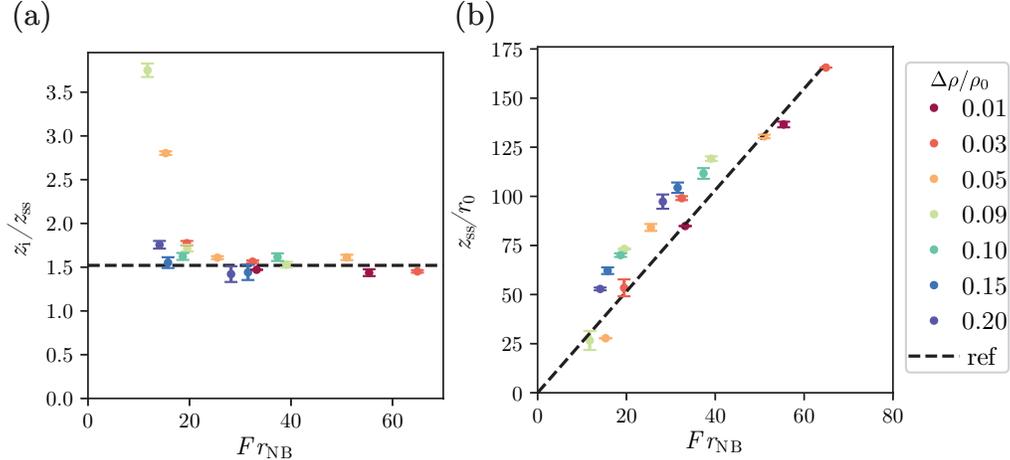}
\caption{(a) The ratio between the initial peak centerline rise height $z_i$ and the time-averaged centerline rise height $z_{ss}$ as a function of non-Boussinesq Froude number $Fr_{\text{NB}}$. The black dashed line represents the reference value 1.52 reported in \citet{Hunt2016}. (b) Steady rise height divided by the needle radius as a function of $Fr_{\text{NB}}$. The black dashed line represents the reference value 2.58 reported in \citet{Mehaddi2015} for a non-Boussinesq round fountain. The error bars measure the standard deviations of the reported values among the repeating experiments.}
\label{img:height}
\end{figure}

To analyze the experiments with saline solutions, we start with the fountain rise height as a function of the Froude number, see Fig.\ \ref{img:height}. Note that we need to subtract the jet laminar length to accurately obtain the rise height. Fig.\ \ref{img:height}(a) shows the height ratio between the initial peak of the rise height and the time-averaged rise height in the steady fountain stage. With a relatively large density difference, we use the non-Boussinesq Froude number \citep{Vaux2019} $Fr_{\text{NB}} = Fr_0(\rho_i/\rho_0)^{0.75}$ to replace the commonly used source Froude number $Fr_0$, which is adapted from \citet{Vaux2019} for upward round fountain. For $Fr_{\text{NB}}>20$, the experimental data agrees well with the reference value 1.52 reported in \citet{Hunt2016} for a round fountain. Fig.\ \ref{img:height}(b) presents the time-averaged rise height scaling with $Fr_{\text{NB}}$ in the steady fountain stage. The rise heights of the fountains with relatively low density differences are well captured by the prediction in \citet{Mehaddi2015} for a non-Boussinesq round fountain, namely $z_{ss}/r_0=2.58Fr_{\text{NB}}$, where $z_{ss}$ is the steady-state fountain height. \citet{Debugne2018} also showed that the rise height scaling for a round fountain also works for a confined fountain. The causes of the deviation from the reference value for the fountains with higher density differences, however, remain unclear. A combined effect of the non-Boussinesq effect and the shear dispersion within a quasi-2D geometry can be a possible explanation. 

\begin{figure}[h!bt]
\centering
\includegraphics[scale=1]{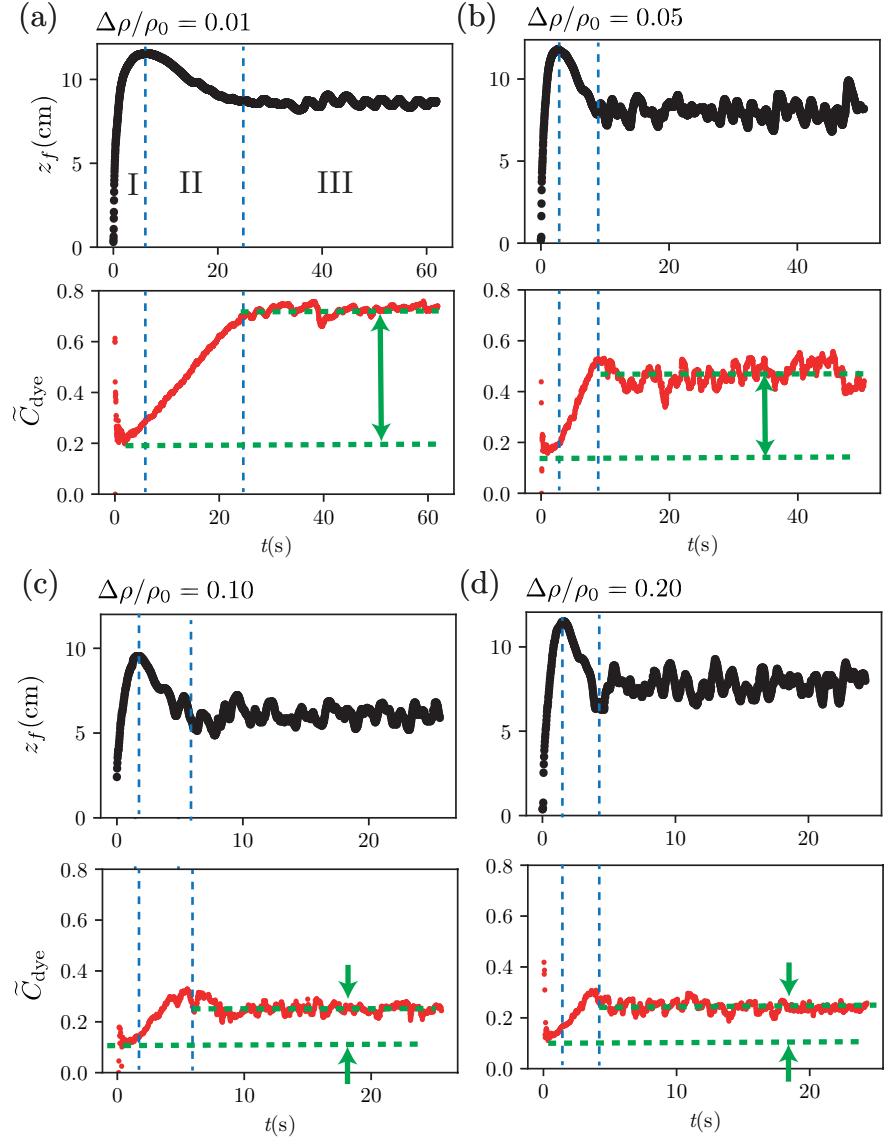}
\caption{Temporal evolution of the fountain rise height and the fountain core concentration for (a) Exp.\ B1, (b) Exp.\ B7, (c) Exp.\ B12, and (d) Exp.\ B17 in Table \ref{tbl:condition4}. The definition of the fountain rise height and the fountain core are the same as that in Fig.\ \ref{img:coreeth}(a). The blue dashed lines divide a fountain into three stages, (I) NBJ, (II) flow reversal, and (III) steady fountain. The green dashed lines and arrows mark the difference of core concentration between the NBJ stage and that in the steady fountain stage.}
\label{img:ccore_evolution}
\end{figure}

To evaluate the degree of mixing in the saline fountains, we now track the temporal evolution of the fountain core concentration. Fig.\ \ref{img:ccore_evolution} presents the evolution of the core concentration together with the fountain top position $z_f$ of several experiments with increasing density difference, see Figs.\ \ref{img:ccore_evolution}(a--d). As mentioned in \S4.3.1, we define three stages of a fountain using the evolution of $z_f$, namely (I) the NBJ stage, (II) the flow reversal, and (III) the (quasi-)steady fountain stage. Note that the start of the regime (III) roughly matches the local maximum of the core concentration, as indicated by the second blue dashed line in Figs.\ \ref{img:ccore_evolution}(a--d). The evolutions of the core concentrations are pretty similar for all the experimental conditions. Upon injection, the core concentrations decrease sharply, indicating intense dilution in the early NBJ stage. In the rest of the NBJ stage and the flow reversal stage, the core concentrations keep increasing until a plateau in the steady fountain stage. The increasing concentrations suggest weakening of the dilution, which we attribute to the shielding of the outer flow in the confined geometry. 

The green arrows in Figs.\ \ref{img:ccore_evolution}(a--d) help us to identify that the difference between the core concentration in the NBJ stage and that in the steady fountain stage decreases with $\Delta \rho/\rho_0$, at least in the range $0.01 \leq \Delta \rho/\rho_0 \leq 0.10$. The reduced difference suggests that the shielding effect of the outer flow decreases with increasing $\Delta \rho/\rho_0$.   

\begin{figure}
\centering
\includegraphics[scale=1]{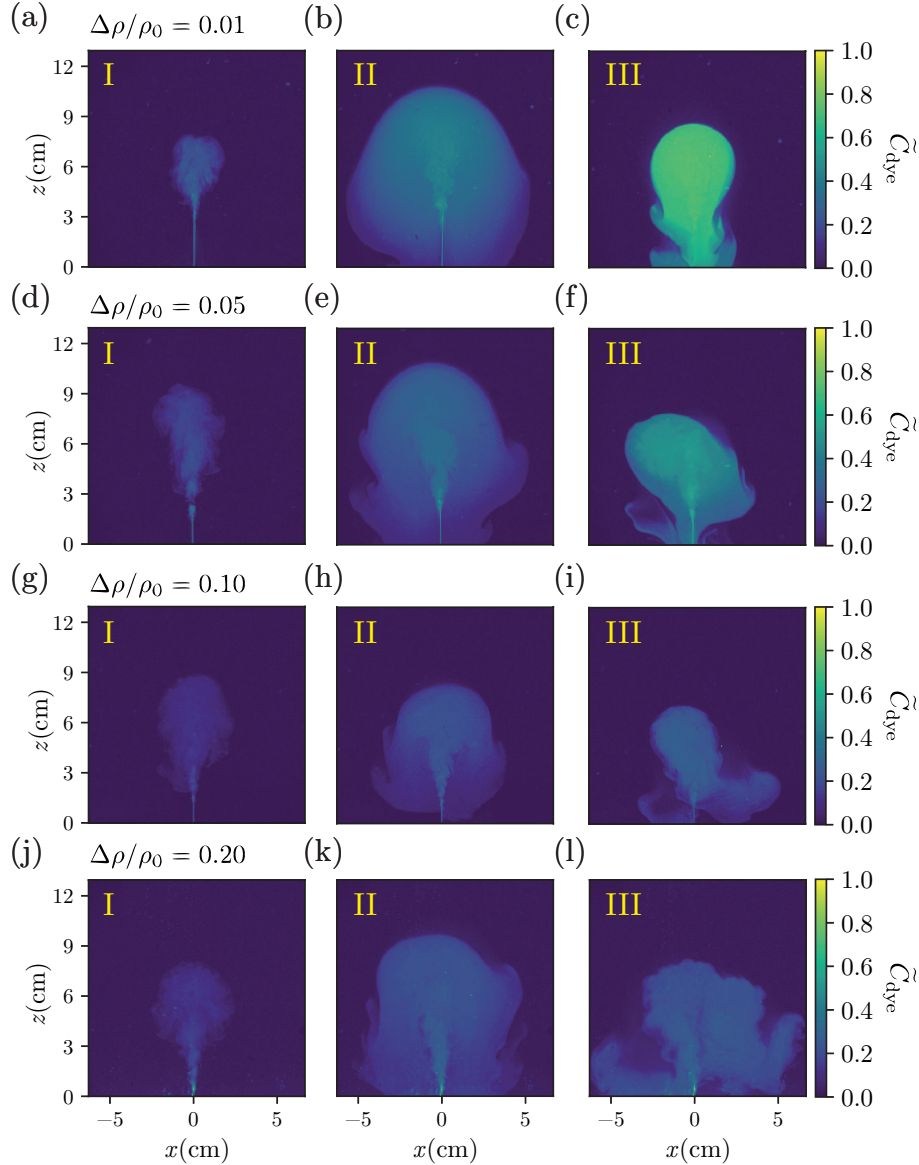}
\caption{Snapshots of the saline fountains for (a--c) Exp.\ B1, (d--f) Exp.\ B7, (g--i) Exp.\ B12, and (j--l) Exp.\ B17 in Table \ref{tbl:condition4}. (a,d,g,j) show the snapshots in the NBJ stage (I), (b,e,h,k) the flow reversal stage (II), and (c,f,i,l) the steady fountain stage (III).}
\label{img:snapshots}
\end{figure}

Fig.\ \ref{img:snapshots} show the snapshots of the three stages for the corresponding saline fountains displayed in Fig.\ \ref{img:ccore_evolution}. The concentration fields in the snapshots agree with the temporal evolution of the core concentrations, exhibiting weakening dilution as the flow goes through the three stages. Fig.\ \ref{img:snapshots} also provides visualizations of the shielding effect we discussed earlier, which is initiated in the flow reversal stage (II). Fig.\ \ref{img:snapshots}(b) clearly shows shielding formed by the outer flow, which reduces the oscillations of the fountain. With a larger $\Delta \rho/\rho_0$, the flow enters the steady fountain stage earlier, and the quasi-steady flapping motions start earlier. The quasi-steady flapping motions of the confined fountains have been discussed in detail in \citet{Debugne2018}, which is a unique feature in the meandering regime $(Fr_0>16)$. All of our experiments are in this regime except for Exps.\ B9 and B14, see Table \ref{tbl:condition4}. Note that the magnitude of the flapping depends on $Re_0$ and $Fr_0$, which are not carefully aligned here and beyond the scope of our study. 

\begin{figure}[h!tb]
\centering
\hspace{-5mm}
\includegraphics[scale=1]{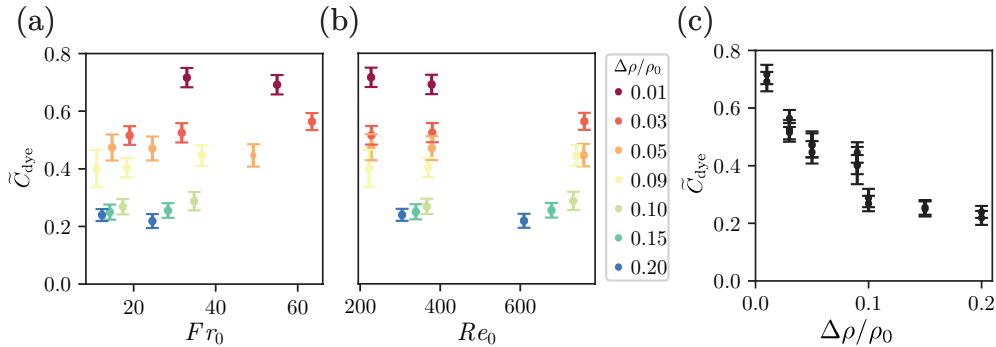}
\caption{Time-averaged core concentration as a function of (a) $Fr_0$, (b) $Re_0$, and (c) $\Delta \rho/\rho_0$. The error bars are the standard deviation of the concentrations.}
\label{img:ccoremean}
\end{figure}

To study the effect of $Fr_0$ and $Re_0$ on mixing, we calculate the time-averaged core concentration in the steady fountain stage for all the saline fountains, Exps.\ B1--B17. Figs.\ \ref{img:ccoremean}(a,b) show that $Fr_0$ and $Re_0$ do not have a major effect on the mean core concentration. Instead, $\Delta \rho/\rho_0$ seems to be the dominant factor. Fig.\ \ref{img:ccoremean}(c) clearly demonstrates that the degree of mixing increases with increasing $\Delta \rho/\rho_0$. Moreover, the degree of mixing seems to approach an asymptote for $\Delta \rho/\rho_0>0.1$, which suggests the existence of an upper bound for mixing in a quasi-2D fountain.   

\begin{figure}[h!tb]
\centering
\hspace{-5mm}
\includegraphics[scale=1]{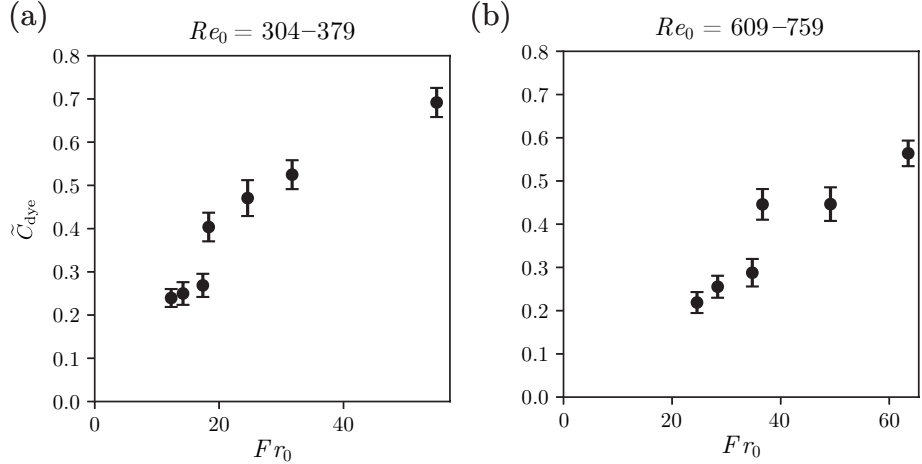}
\caption{Time-averaged core concentration as a function of $Fr_0$ for (a) $Re_0$=304--379, and (b) $Re_0$=609--759.}
\label{img:ccore_fixRe}
\end{figure}

While $Fr_0$ strongly depends on $\Delta \rho/\rho_0$, we extract the mean core concentrations in Figs.\ \ref{img:ccoremean}(a,b) with similar $Re_0$, grouping them into a lower $Re_0$ group and a higher $Re_0$ one, see Figs.\ \ref{img:ccore_fixRe}(a) and (b), respectively. Figs.\ \ref{img:ccore_fixRe}(a) and (b) convey the same message, that is, with a similar magnitude of momentum, a larger $\Delta \rho/\rho_0$ indicates a stronger buoyancy effect, and in turn a smaller $Fr_0$. A stronger buoyancy leads to a weaker shielding effect and more intense flapping motions, as illustrated in Fig.\ \ref{img:snapshots}. Therefore, when $Re_0$ is more or less fixed, the degree of mixing in a quasi-2D fountain decreases with $Fr_0$.    

\begin{figure}[h!tb]
\centering
\includegraphics[scale=1]{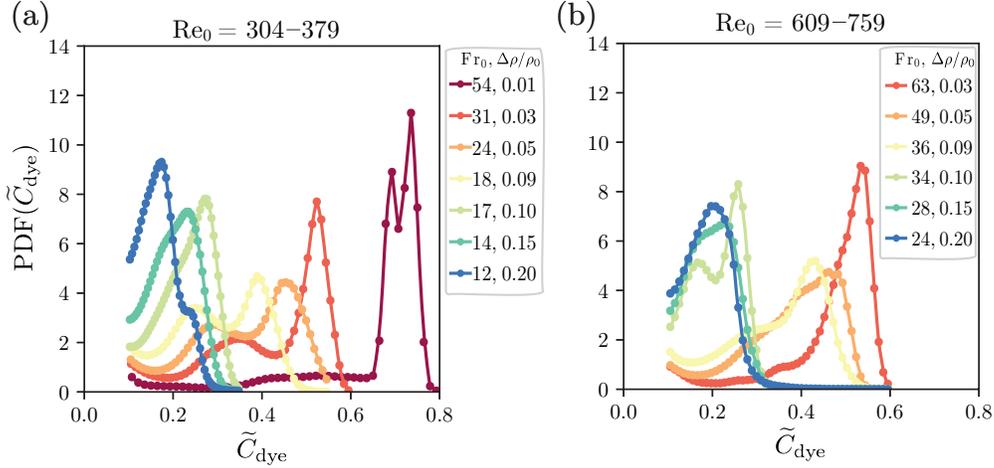}
\caption{Probability density functions (PDF) of the concentration of the entire fountain within the steady state. (a) For $Re_0$=304--379, and (b) for $Re_0$=609--759. In the legends, we show $Fr_0$ and the corresponding $\Delta \rho/\rho_0$, which show the opposite trends of variation.}
\label{img:pdf_fixRe}
\end{figure}

Although the core concentration sheds light on mixing in a quasi-2D fountain, it might not be representative enough to reach a conclusion about mixing in the entire fountain. Consequently, we calculate the probability density functions (PDFs) of the entire fountain in the steady fountain stage for all the saline experiments. The edge of the fountain is detected using a threshold value determined from the histogram of the concentration fields, similar to what we did in Chapter 2. Then we group the PDFs with similar $Re_0$ into a lower $Re_0$ group in Fig.\ \ref{img:pdf_fixRe}(a) and a higher $Re_0$ group in Fig.\ \ref{img:pdf_fixRe}(b). For the lower $Re_0$ group in Fig.\ \ref{img:pdf_fixRe}(a), the spread between the profiles is more pronounced, suggesting a monotonic effect of buoyancy on mixing when the momentum is low. Fig.\ \ref{img:pdf_fixRe}(b), on the other hand, shows that the variation of the profiles is less continuous, exhibiting three groups with two major shifts located at $49\leq Fr_0 \leq63$ and $34\leq Fr_0 \leq36$. The two critical $Fr_0$s divide the flow into three mixing regimes. Away from the two critical $Fr_0$s, the effect of $Fr_0$ on mixing within each regime is minor.   

The profile in Fig.\ \ref{img:pdf_fixRe}(b) is skewed more towards the high concentration side than the corresponding case in Fig.\ \ref{img:pdf_fixRe}(a) (with the same $\Delta \rho/\rho_0$). Under the same saline composition ($\Delta \rho/\rho_0$), $Fr_0$ and $Re_0$ increase in the same way when we raise the flow rate, which lowers the degree of mixing in the whole fountain due to the reduced effect of buoyancy. As this dependence is not clearly revealed in the mean core concentration shown in Figs.\ \ref{img:ccoremean}(a--b), it confirms the importance to look at the PDFs of the concentration when evaluating mixing in a quasi-2D fountain. 

\begin{figure}[h!tb]
\centering
\includegraphics[scale=1]{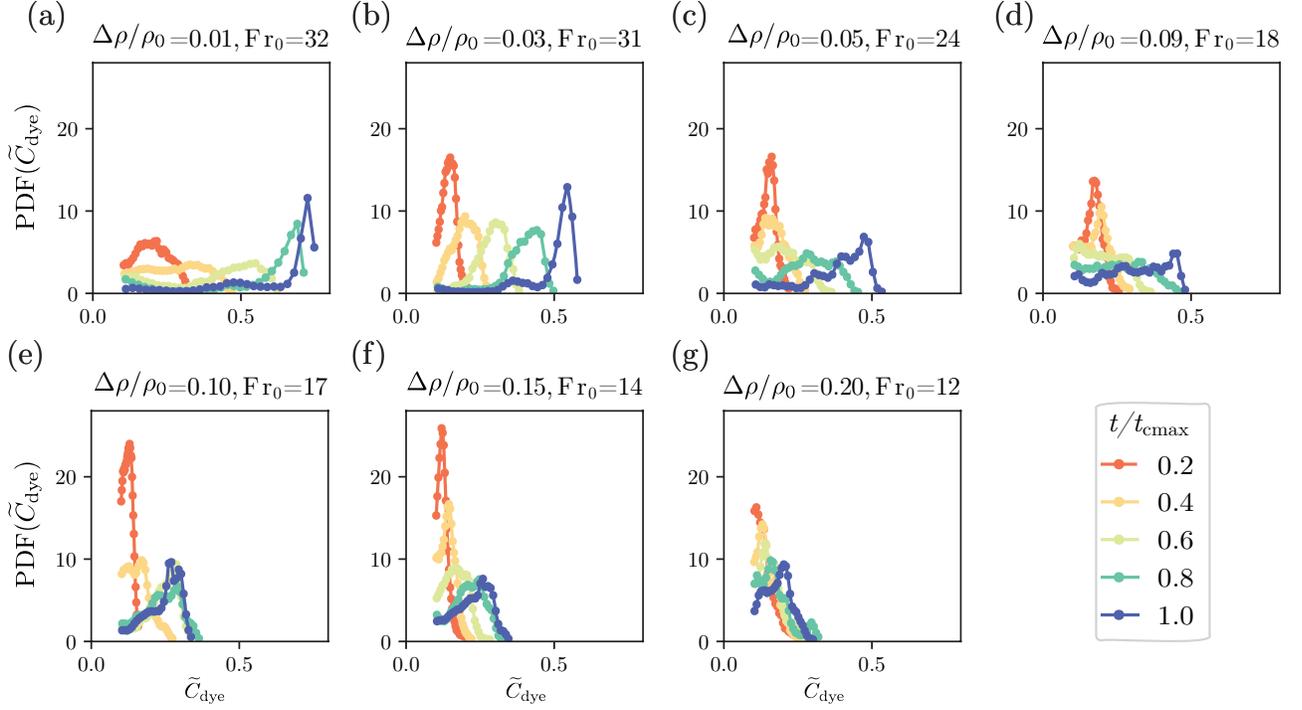}
\caption{Probability density functions (PDFs) of the concentration in the NBJ stage for (a) Exp.\ B1, (b) Exp.\ B4, (c) Exp.\ B7, (d) Exp.\ B10, (e) Exp.\ B12, (f) Exp.\ B14, and (g) Exp.\ B16 in Table \ref{tbl:condition4}. $t_{\text{cmax}}$ in the legend denotes the start of the steady fountain stage (III).}
\label{img:pdf_nbj}
\end{figure}

\begin{figure}[h!tb]
\centering
\includegraphics[scale=1]{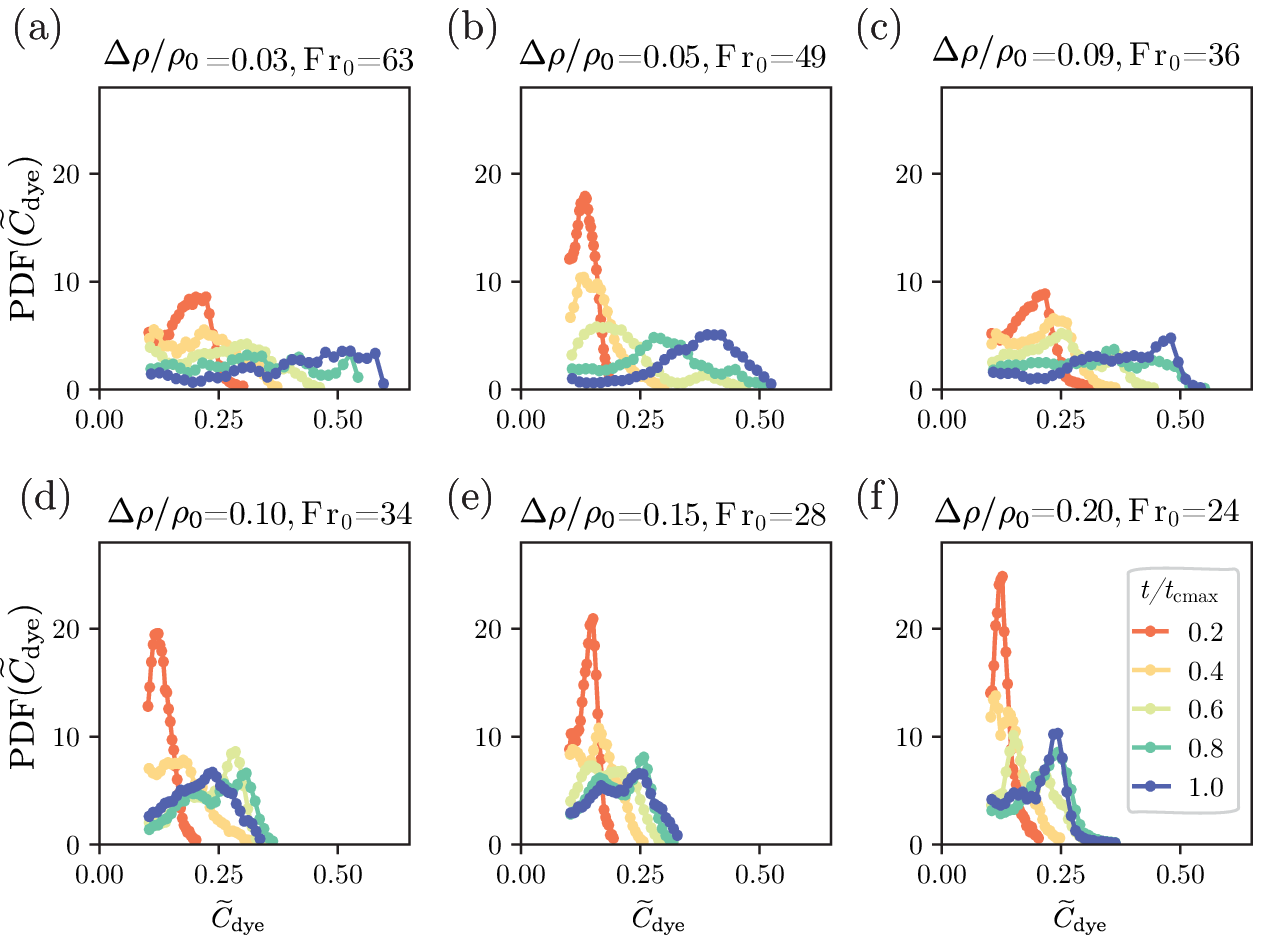}
\caption{Probability density functions (PDFs) of the concentration in the NBJ stage for (a) Exp.\ B5, (b) Exp.\ B8, (c) Exp.\ B11, (d) Exp.\ B13, (e) Exp.\ B15, and (f) Exp.\ B17 in Table \ref{tbl:condition4}.}
\label{img:pdf_nbj2}
\end{figure}

In \S4.3.1 we revealed that mixing in the steady fountain stage (III) is largely determined by that in the NBJ and the flow reversal stages. Despite the difference in the direction of injection and the needle length for the upward saline fountain, we expect that the state of mixing in the steady fountain stage can be traced back to that in the prior stages. Therefore, we plot the PDFs of the concentrations before the steady fountain stage in Figs.\ \ref{img:pdf_nbj} and \ref{img:pdf_nbj2}. Each subfigure in Fig.\ \ref{img:pdf_nbj} finds its corresponding profile in Fig.\ \ref{img:pdf_fixRe}(a), and each subfigure in Fig.\ \ref{img:pdf_nbj2} finds its corresponding profile in Fig.\ \ref{img:pdf_fixRe}(b). 

Fig.\ \ref{img:pdf_nbj} reveals that the concentrations gradually evolve to higher values as the flow approaches the steady fountain regime. Moreover, this evolution gets less pronounced with decreasing $Fr_0$ (increasing $\Delta \rho/\rho_0$), which agrees with the evolution of the core concentration presented in Fig.\ \ref{img:ccore_evolution}. Comparing the profiles at the same $t/t_{\text{cmax}}$ across Figs.\ \ref{img:pdf_nbj}(a--g), we can identify that the spread of the profiles with varying $Fr_0$, except for $t/t_{\text{cmax}}=0.2$, aligns with the monotonic spread discovered in Fig.\ \ref{img:pdf_fixRe}. 

In the group with higher $Re_0$ shown in Fig.\ \ref{img:pdf_nbj2}, most of the concentration distributions are similar to the corresponding case in Fig.\ \ref{img:pdf_nbj} (with the same $\Delta \rho/\rho_0$). However, the cases with $\Delta \rho/\rho_0=0.03$ have a noticeable difference. The wider distributions in Fig.\ \ref{img:pdf_nbj2}(a) as compared to Fig.\ \ref{img:pdf_nbj}(b) demonstrate enhanced mixing in the early NBJ stage with larger momentum. Furthermore, comparing the profiles at the same $t/t_{\text{cmax}}$ across Figs.\ \ref{img:pdf_nbj2}(a--f), we can identify the sharp transitions located at $49\leq Fr_0 \leq63$ and $34\leq Fr_0 \leq36$, which agrees with the finding for the steady fountain in Fig.\ \ref{img:pdf_fixRe}(b). 

Figs.\ \ref{img:pdf_fixRe}--\ref{img:pdf_nbj2} show that the effect of $Fr_0$ ($\Delta \rho/\rho_0$) on the PDFs of the concentrations in the steady fountain stage closely follows the concentrations in the NBJ and the flow reversal stages, proving once again that mixing in a highly confined fountain is determined in its developing stages prior to the final quasi-steady state. 
\clearpage
\section{Conclusion}
We have investigated mixing in highly confined fountains using two sets of experiments. In the first set of experiments, we found that the downward fountain formed by an ouzo mixture hardly induce nucleation. The long injection needle causes a laminar NBJ stage and a strong shielding effect initiated in the flow reversal stage, which further reduces the entrainment of ambient water into the inner zone of the fountain. Without sufficient entrainment, mixing and nucleation are restricted to the outer rim of the fountain. The reference experiments using dyed ethanol show that the ethanol fraction remains above 0.6 almost within the entire fountain body, which indeed cannot induce nucleation based on the phase diagram. 

In the second set of experiments, we systematically studied the effect of $\Delta \rho/\rho_0$ on mixing in the highly confined fountains. Saline solutions with different compositions were injected upward, forming fountains with various $\Delta \rho/\rho_0$, $Fr_0$, and $Re_0$. By tracking the temporal evolution of the self-defined core concentration, we identify how the shielding effect reduces the degree of mixing, which decreases from the NBJ stage to the flow reversal stage, and finally reaches a plateau in the quasi-steady fountain stage. Comparing the plateau values in all the saline experiments, we reveal that $\Delta \rho/\rho_0$, instead of $Fr_0$ or $Re_0$, is the critical parameter determining the degree of mixing. Confinement facilitates the shielding effect, which in turn attenuates the stretching process \cite{Villermaux2019}. The larger $\Delta \rho/\rho_0$, the larger the (negative) buoyancy force. The buoyancy force overcomes the shielding effect and enhances the stretching, leading to stronger mixing within the fountain.

To characterize aspects of mixing missing in the core concentration, the PDFs of the concentration in the entire fountain are calculated. The profiles in the low $Re_0$ group show a monotonic decrease of mixing with increasing $Fr_0$ ($\Delta \rho/\rho_0$), while those in the high $Re_0$ group indicate the presence of three mixing regimes divided by two critical $Fr_0$s. The difference in the two groups suggests an effect of $Re_0$ on the degree of mixing. 

Although the two sets of experiments can't be well matched due to a different source condition (the needle length), they both lead to the same conclusion that the degree of mixing in a highly confined fountain is largely determined in its developing stages, where the competition between the buoyancy effect and the shielding effect takes place. 

\clearpage
\section*{Appendix A: Calibration curves}
\begin{figure}[h!tb]
\centering
\includegraphics[scale=1]{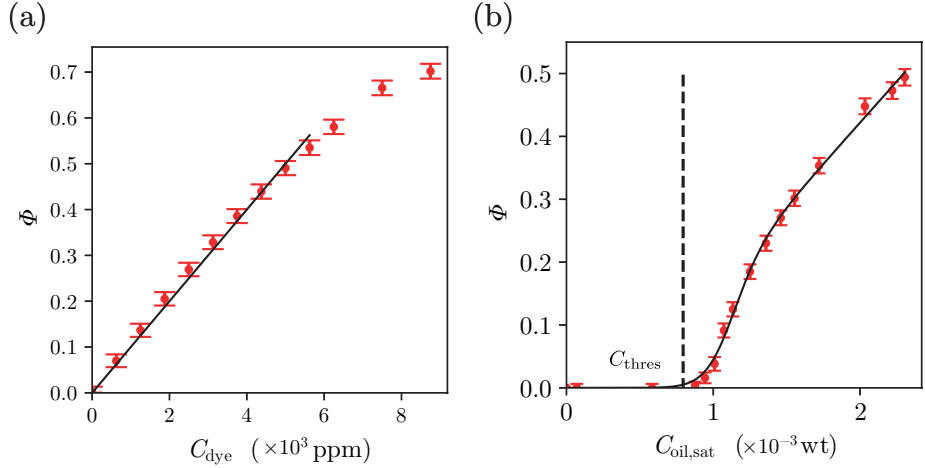}
\caption{The calibration curve for (a) the dyed ethanol and (b) the ouzo mixture. $\Phi$ is the degree of light attenuation, $\log(I_{\text{ref}}/I)$. The red points are the measured data from the calibration, the black curves are the fitted calibration curves, and the black error bars represent the standard deviation of $\Phi$, which are small throughout the calibration. The abscissa $C_{\text{oil,sat}}$ in (b) is the oversaturation of the oil.} 
\label{img:calicurve4}
\end{figure}

We have used in-situ calibration to obtain calibration curves to convert the recorded light attenuation to either dye concentration or oversaturation of the nucleated oil. For the dye case, we obtained calibration data points shown in Fig.\ \ref{img:calicurve4}(a), which can be fitted with a linear curve for concentration below \SI{4000}ppm. For the ouzo case, the calibration data points are obtained using the phase diagram discussed in \S4.2.2 and Fig. \ref{img:phasegram}, which we fit by the empirical function,

\begin{equation}
    \Phi(C) = \log \left( I_{\text{ref}}/I \right)= \frac{a_0}{1+a_1e^{a_2(C-a_3)}}+\frac{a_4(C-a_5)}{1+a_1e^{a_2(C-a_3)}},
\label{eq:LA4}
\end{equation}
where $\Phi$ is the light attenuation level, $C$ the oil concentration (or oversaturation), and $a_1$--$a_5$ fitting parameters.

We define $C_{\text{thres}}$ to be the point where $\Phi$ reaches 1\% of its peak. Note that the calibration curve shown here is a local calibration curve for one image unit of 5$\times$5 pixels. We have obtained in total 204 $\times$ 204 local calibration curves within the recording domain.

\begin{acknowledgments}
The authors acknowledge the funding by ERC Advanced Grant Diffusive Droplet Dynamics (DDD) with Project No. 740479 and Netherlands Organisation for Scientific Research (NWO) through the Multiscale Catalytic Energy Conversion (MCEC) research center.

We thank G.W. Bruggert, M. Bos, and T. Zijlstra for technical support in building the setup. 
\end{acknowledgments}

\bibliography{PRF-fountain}

\begin{thebibliography}{31}%
\makeatletter
\providecommand \@ifxundefined [1]{%
 \@ifx{#1\undefined}
}%
\providecommand \@ifnum [1]{%
 \ifnum #1\expandafter \@firstoftwo
 \else \expandafter \@secondoftwo
 \fi
}%
\providecommand \@ifx [1]{%
 \ifx #1\expandafter \@firstoftwo
 \else \expandafter \@secondoftwo
 \fi
}%
\providecommand \natexlab [1]{#1}%
\providecommand \enquote  [1]{``#1''}%
\providecommand \bibnamefont  [1]{#1}%
\providecommand \bibfnamefont [1]{#1}%
\providecommand \citenamefont [1]{#1}%
\providecommand \href@noop [0]{\@secondoftwo}%
\providecommand \href [0]{\begingroup \@sanitize@url \@href}%
\providecommand \@href[1]{\@@startlink{#1}\@@href}%
\providecommand \@@href[1]{\endgroup#1\@@endlink}%
\providecommand \@sanitize@url [0]{\catcode `\\12\catcode `\$12\catcode
  `\&12\catcode `\#12\catcode `\^12\catcode `\_12\catcode `\%12\relax}%
\providecommand \@@startlink[1]{}%
\providecommand \@@endlink[0]{}%
\providecommand \url  [0]{\begingroup\@sanitize@url \@url }%
\providecommand \@url [1]{\endgroup\@href {#1}{\urlprefix }}%
\providecommand \urlprefix  [0]{URL }%
\providecommand \Eprint [0]{\href }%
\providecommand \doibase [0]{https://doi.org/}%
\providecommand \selectlanguage [0]{\@gobble}%
\providecommand \bibinfo  [0]{\@secondoftwo}%
\providecommand \bibfield  [0]{\@secondoftwo}%
\providecommand \translation [1]{[#1]}%
\providecommand \BibitemOpen [0]{}%
\providecommand \bibitemStop [0]{}%
\providecommand \bibitemNoStop [0]{.\EOS\space}%
\providecommand \EOS [0]{\spacefactor3000\relax}%
\providecommand \BibitemShut  [1]{\csname bibitem#1\endcsname}%
\let\auto@bib@innerbib\@empty
\bibitem [{\citenamefont {Villermaux}(2019)}]{Villermaux2019}%
  \BibitemOpen
  \bibfield  {author} {\bibinfo {author} {\bibfnamefont {E.}~\bibnamefont
  {Villermaux}},\ }\bibfield  {title} {\bibinfo {title} {Mixing versus
  stirring},\ }\href@noop {} {\bibfield  {journal} {\bibinfo  {journal} {Annu.
  Rev. Fluid Mech.}\ }\textbf {\bibinfo {volume} {51}},\ \bibinfo {pages} {245}
  (\bibinfo {year} {2019})}\BibitemShut {NoStop}%
\bibitem [{\citenamefont {Hunt}\ and\ \citenamefont
  {Burridge}(2015)}]{Hunt2015}%
  \BibitemOpen
  \bibfield  {author} {\bibinfo {author} {\bibfnamefont {G.}~\bibnamefont
  {Hunt}}\ and\ \bibinfo {author} {\bibfnamefont {H.}~\bibnamefont
  {Burridge}},\ }\bibfield  {title} {\bibinfo {title} {Fountains in industry
  and nature},\ }\href@noop {} {\bibfield  {journal} {\bibinfo  {journal}
  {Annu. Rev. Fluid Mech.}\ }\textbf {\bibinfo {volume} {47}},\ \bibinfo
  {pages} {195–220} (\bibinfo {year} {2015})}\BibitemShut {NoStop}%
\bibitem [{\citenamefont {Williamson}\ \emph {et~al.}(2011)\citenamefont
  {Williamson}, \citenamefont {Armfield},\ and\ \citenamefont
  {Lin}}]{Williamson2011}%
  \BibitemOpen
  \bibfield  {author} {\bibinfo {author} {\bibfnamefont {N.}~\bibnamefont
  {Williamson}}, \bibinfo {author} {\bibfnamefont {S.}~\bibnamefont
  {Armfield}},\ and\ \bibinfo {author} {\bibfnamefont {W.}~\bibnamefont
  {Lin}},\ }\bibfield  {title} {\bibinfo {title} {Forced turbulent fountain
  flow behaviour},\ }\href@noop {} {\bibfield  {journal} {\bibinfo  {journal}
  {J. Fluid Mech.}\ }\textbf {\bibinfo {volume} {671}},\ \bibinfo {pages} {535}
  (\bibinfo {year} {2011})}\BibitemShut {NoStop}%
\bibitem [{\citenamefont {Burridge}\ and\ \citenamefont
  {Hunt}(2012)}]{Burridge2012}%
  \BibitemOpen
  \bibfield  {author} {\bibinfo {author} {\bibfnamefont {H.}~\bibnamefont
  {Burridge}}\ and\ \bibinfo {author} {\bibfnamefont {G.}~\bibnamefont
  {Hunt}},\ }\bibfield  {title} {\bibinfo {title} {The rise heights of low- and
  high-{F}roude-number turbulent axisymmetric fountains},\ }\href@noop {}
  {\bibfield  {journal} {\bibinfo  {journal} {J. Fluid Mech.}\ }\textbf
  {\bibinfo {volume} {691}},\ \bibinfo {pages} {392} (\bibinfo {year}
  {2012})}\BibitemShut {NoStop}%
\bibitem [{\citenamefont {Burridge}\ and\ \citenamefont
  {Hunt}(2013)}]{Burridge2013}%
  \BibitemOpen
  \bibfield  {author} {\bibinfo {author} {\bibfnamefont {H.}~\bibnamefont
  {Burridge}}\ and\ \bibinfo {author} {\bibfnamefont {G.}~\bibnamefont
  {Hunt}},\ }\bibfield  {title} {\bibinfo {title} {The rhythm of fountains: The
  length and time scales of rise height fluctuations at low and high {F}roude
  numbers},\ }\href@noop {} {\bibfield  {journal} {\bibinfo  {journal} {J.
  Fluid Mech.}\ }\textbf {\bibinfo {volume} {728}},\ \bibinfo {pages} {91}
  (\bibinfo {year} {2013})}\BibitemShut {NoStop}%
\bibitem [{\citenamefont {Burridge}\ and\ \citenamefont
  {Hunt}(2016)}]{Burridge2016}%
  \BibitemOpen
  \bibfield  {author} {\bibinfo {author} {\bibfnamefont {H.}~\bibnamefont
  {Burridge}}\ and\ \bibinfo {author} {\bibfnamefont {G.}~\bibnamefont
  {Hunt}},\ }\bibfield  {title} {\bibinfo {title} {Entrainment by turbulent
  fountains},\ }\href@noop {} {\bibfield  {journal} {\bibinfo  {journal} {J.
  Fluid Mech.}\ }\textbf {\bibinfo {volume} {790}},\ \bibinfo {pages} {407}
  (\bibinfo {year} {2016})}\BibitemShut {NoStop}%
\bibitem [{\citenamefont {Milton-McGurk}\ \emph {et~al.}(2022)\citenamefont
  {Milton-McGurk}, \citenamefont {Williamson}, \citenamefont {Armfield},\ and\
  \citenamefont {Kirkpatrick}}]{Milton2022}%
  \BibitemOpen
  \bibfield  {author} {\bibinfo {author} {\bibfnamefont {L.}~\bibnamefont
  {Milton-McGurk}}, \bibinfo {author} {\bibfnamefont {N.}~\bibnamefont
  {Williamson}}, \bibinfo {author} {\bibfnamefont {S.}~\bibnamefont
  {Armfield}},\ and\ \bibinfo {author} {\bibfnamefont {M.}~\bibnamefont
  {Kirkpatrick}},\ }\bibfield  {title} {\bibinfo {title} {Characterising
  entrainment in fountains and negatively buoyant jets},\ }\href@noop {}
  {\bibfield  {journal} {\bibinfo  {journal} {J. Fluid Mech.}\ }\textbf
  {\bibinfo {volume} {939}},\ \bibinfo {pages} {A29} (\bibinfo {year}
  {2022})}\BibitemShut {NoStop}%
\bibitem [{\citenamefont {Talluru}\ \emph {et~al.}(2022)\citenamefont
  {Talluru}, \citenamefont {Williamson},\ and\ \citenamefont
  {Armfield}}]{Talluru2022}%
  \BibitemOpen
  \bibfield  {author} {\bibinfo {author} {\bibfnamefont {K.}~\bibnamefont
  {Talluru}}, \bibinfo {author} {\bibfnamefont {N.}~\bibnamefont
  {Williamson}},\ and\ \bibinfo {author} {\bibfnamefont {S.}~\bibnamefont
  {Armfield}},\ }\bibfield  {title} {\bibinfo {title} {Entrainment and dilution
  in a fountain top},\ }\href@noop {} {\bibfield  {journal} {\bibinfo
  {journal} {J. Fluid Mech.}\ }\textbf {\bibinfo {volume} {941}},\ \bibinfo
  {pages} {A24} (\bibinfo {year} {2022})}\BibitemShut {NoStop}%
\bibitem [{\citenamefont {Xue}\ \emph {et~al.}(2019)\citenamefont {Xue},
  \citenamefont {Khodaparast},\ and\ \citenamefont {Stone}}]{Xue2019}%
  \BibitemOpen
  \bibfield  {author} {\bibinfo {author} {\bibfnamefont {N.}~\bibnamefont
  {Xue}}, \bibinfo {author} {\bibfnamefont {S.}~\bibnamefont {Khodaparast}},\
  and\ \bibinfo {author} {\bibfnamefont {H.~A.}\ \bibnamefont {Stone}},\
  }\bibfield  {title} {\bibinfo {title} {Fountain mixing in a filling box at
  low {R}eynolds numbers},\ }\href@noop {} {\bibfield  {journal} {\bibinfo
  {journal} {Phys. Rev. Fluids}\ }\textbf {\bibinfo {volume} {4}},\ \bibinfo
  {pages} {024501} (\bibinfo {year} {2019})}\BibitemShut {NoStop}%
\bibitem [{\citenamefont {Baines}\ and\ \citenamefont
  {Turner}(1969)}]{Baines1969}%
  \BibitemOpen
  \bibfield  {author} {\bibinfo {author} {\bibfnamefont {W.~D.}\ \bibnamefont
  {Baines}}\ and\ \bibinfo {author} {\bibfnamefont {J.~S.}\ \bibnamefont
  {Turner}},\ }\bibfield  {title} {\bibinfo {title} {Turbulent buoyant
  convection from a source in a confined region},\ }\href@noop {} {\bibfield
  {journal} {\bibinfo  {journal} {J. Fluid Mech.}\ }\textbf {\bibinfo {volume}
  {37}},\ \bibinfo {pages} {51} (\bibinfo {year} {1969})}\BibitemShut {NoStop}%
\bibitem [{\citenamefont {Hunt}\ \emph {et~al.}(2019)\citenamefont {Hunt},
  \citenamefont {Debugne},\ and\ \citenamefont {Ciriello}}]{Hunt2019}%
  \BibitemOpen
  \bibfield  {author} {\bibinfo {author} {\bibfnamefont {G.}~\bibnamefont
  {Hunt}}, \bibinfo {author} {\bibfnamefont {A.}~\bibnamefont {Debugne}},\ and\
  \bibinfo {author} {\bibfnamefont {F.}~\bibnamefont {Ciriello}},\ }\bibfield
  {title} {\bibinfo {title} {The structure of a turbulent line fountain},\
  }\href@noop {} {\bibfield  {journal} {\bibinfo  {journal} {J. Fluid Mech.}\
  }\textbf {\bibinfo {volume} {876}},\ \bibinfo {pages} {680} (\bibinfo {year}
  {2019})}\BibitemShut {NoStop}%
\bibitem [{\citenamefont {Debugne}\ and\ \citenamefont
  {Hunt}(2018)}]{Debugne2018}%
  \BibitemOpen
  \bibfield  {author} {\bibinfo {author} {\bibfnamefont {A.}~\bibnamefont
  {Debugne}}\ and\ \bibinfo {author} {\bibfnamefont {G.}~\bibnamefont {Hunt}},\
  }\bibfield  {title} {\bibinfo {title} {The influence of spanwise confinement
  on round fountains},\ }\href@noop {} {\bibfield  {journal} {\bibinfo
  {journal} {J. Fluid Mech.}\ }\textbf {\bibinfo {volume} {845}},\ \bibinfo
  {pages} {263} (\bibinfo {year} {2018})}\BibitemShut {NoStop}%
\bibitem [{\citenamefont {Rothstein}\ \emph {et~al.}(1999)\citenamefont
  {Rothstein}, \citenamefont {Henry},\ and\ \citenamefont
  {Gollub}}]{Rothstein1999}%
  \BibitemOpen
  \bibfield  {author} {\bibinfo {author} {\bibfnamefont {D.}~\bibnamefont
  {Rothstein}}, \bibinfo {author} {\bibfnamefont {E.}~\bibnamefont {Henry}},\
  and\ \bibinfo {author} {\bibfnamefont {J.}~\bibnamefont {Gollub}},\
  }\bibfield  {title} {\bibinfo {title} {Persistent patterns in transient
  chaotic fluid mixing},\ }\href@noop {} {\bibfield  {journal} {\bibinfo
  {journal} {Nature}\ }\textbf {\bibinfo {volume} {401}},\ \bibinfo {pages}
  {770} (\bibinfo {year} {1999})}\BibitemShut {NoStop}%
\bibitem [{\citenamefont {Duplat}\ and\ \citenamefont
  {VILLERMAUX}(2008)}]{Duplat2008}%
  \BibitemOpen
  \bibfield  {author} {\bibinfo {author} {\bibfnamefont {J.}~\bibnamefont
  {Duplat}}\ and\ \bibinfo {author} {\bibfnamefont {E.}~\bibnamefont
  {VILLERMAUX}},\ }\bibfield  {title} {\bibinfo {title} {Mixing by random
  stirring in confined mixtures},\ }\href@noop {} {\bibfield  {journal}
  {\bibinfo  {journal} {J. Fluid Mech.}\ }\textbf {\bibinfo {volume} {617}},\
  \bibinfo {pages} {51} (\bibinfo {year} {2008})}\BibitemShut {NoStop}%
\bibitem [{\citenamefont {Batchelor.}(1959)}]{Batchelor1959}%
  \BibitemOpen
  \bibfield  {author} {\bibinfo {author} {\bibfnamefont {G.}~\bibnamefont
  {Batchelor.}},\ }\bibfield  {title} {\bibinfo {title} {Small-scale variation
  of convected quantities like temperature in turbulent fluid. part 1. general
  discussion and the case of small conductivity},\ }\href@noop {} {\bibfield
  {journal} {\bibinfo  {journal} {J. Fluid Mech.}\ }\textbf {\bibinfo {volume}
  {5}},\ \bibinfo {pages} {113–133} (\bibinfo {year} {1959})}\BibitemShut
  {NoStop}%
\bibitem [{\citenamefont {Mingotti}\ and\ \citenamefont
  {Cardoso}(2019)}]{Mingotti2019b}%
  \BibitemOpen
  \bibfield  {author} {\bibinfo {author} {\bibfnamefont {N.}~\bibnamefont
  {Mingotti}}\ and\ \bibinfo {author} {\bibfnamefont {S.}~\bibnamefont
  {Cardoso}},\ }\bibfield  {title} {\bibinfo {title} {Mixing and reaction in
  turbulent plumes: The limits of slow and instantaneous chemical kinetics},\
  }\href@noop {} {\bibfield  {journal} {\bibinfo  {journal} {J. Fluid Mech.}\
  }\textbf {\bibinfo {volume} {861}},\ \bibinfo {pages} {1} (\bibinfo {year}
  {2019})}\BibitemShut {NoStop}%
\bibitem [{\citenamefont {Guilbert}\ and\ \citenamefont
  {Villermaux}(2021)}]{Guilbert2021a}%
  \BibitemOpen
  \bibfield  {author} {\bibinfo {author} {\bibfnamefont {E.}~\bibnamefont
  {Guilbert}}\ and\ \bibinfo {author} {\bibfnamefont {E.}~\bibnamefont
  {Villermaux}},\ }\bibfield  {title} {\bibinfo {title} {Chemical reactions
  rectify mixtures composition},\ }\href@noop {} {\bibfield  {journal}
  {\bibinfo  {journal} {Phys. Rev. Fluids}\ }\textbf {\bibinfo {volume} {6}},\
  \bibinfo {pages} {L112501} (\bibinfo {year} {2021})}\BibitemShut {NoStop}%
\bibitem [{\citenamefont {Guilbert}\ \emph {et~al.}(2021)\citenamefont
  {Guilbert}, \citenamefont {Almarcha},\ and\ \citenamefont
  {Villermaux}}]{Guilbert2021b}%
  \BibitemOpen
  \bibfield  {author} {\bibinfo {author} {\bibfnamefont {E.}~\bibnamefont
  {Guilbert}}, \bibinfo {author} {\bibfnamefont {C.}~\bibnamefont {Almarcha}},\
  and\ \bibinfo {author} {\bibfnamefont {E.}~\bibnamefont {Villermaux}},\
  }\bibfield  {title} {\bibinfo {title} {Chemical reaction for mixing
  studies},\ }\href@noop {} {\bibfield  {journal} {\bibinfo  {journal} {Phys.
  Rev. Fluids}\ }\textbf {\bibinfo {volume} {6}},\ \bibinfo {pages} {114501}
  (\bibinfo {year} {2021})}\BibitemShut {NoStop}%
\bibitem [{\citenamefont {Lee}\ \emph {et~al.}(2022)\citenamefont {Lee},
  \citenamefont {Sun},\ and\ \citenamefont {Lohse}}]{Lee2022}%
  \BibitemOpen
  \bibfield  {author} {\bibinfo {author} {\bibfnamefont {Y.}~\bibnamefont
  {Lee}}, \bibinfo {author} {\bibfnamefont {S.}~\bibnamefont {Sun},
  \bibfnamefont {C.and~Huisman}},\ and\ \bibinfo {author} {\bibfnamefont
  {D.}~\bibnamefont {Lohse}},\ }\bibfield  {title} {\bibinfo {title}
  {Micro-droplet nucleation through solvent exchange in a turbulent buoyant
  jet},\ }\href@noop {} {\bibfield  {journal} {\bibinfo  {journal} {J. Fluid
  Mech.}\ }\textbf {\bibinfo {volume} {943}},\ \bibinfo {pages} {A11} (\bibinfo
  {year} {2022})}\BibitemShut {NoStop}%
\bibitem [{\citenamefont {Zhang}\ \emph {et~al.}(2015)\citenamefont {Zhang},
  \citenamefont {Lu}, \citenamefont {Tan}, \citenamefont {Bao}, \citenamefont
  {He}, \citenamefont {Sun},\ and\ \citenamefont {Lohse}}]{Zhang2015}%
  \BibitemOpen
  \bibfield  {author} {\bibinfo {author} {\bibfnamefont {X.}~\bibnamefont
  {Zhang}}, \bibinfo {author} {\bibfnamefont {Z.}~\bibnamefont {Lu}}, \bibinfo
  {author} {\bibfnamefont {H.}~\bibnamefont {Tan}}, \bibinfo {author}
  {\bibfnamefont {L.}~\bibnamefont {Bao}}, \bibinfo {author} {\bibfnamefont
  {Y.}~\bibnamefont {He}}, \bibinfo {author} {\bibfnamefont {C.}~\bibnamefont
  {Sun}},\ and\ \bibinfo {author} {\bibfnamefont {D.}~\bibnamefont {Lohse}},\
  }\bibfield  {title} {\bibinfo {title} {Formation of surface nanodroplets
  under controlled flow conditions},\ }\href@noop {} {\bibfield  {journal}
  {\bibinfo  {journal} {PNAS}\ }\textbf {\bibinfo {volume} {112}},\ \bibinfo
  {pages} {9253} (\bibinfo {year} {2015})}\BibitemShut {NoStop}%
\bibitem [{\citenamefont {Hajian}\ and\ \citenamefont
  {Hardt}(2015)}]{Hajian2015}%
  \BibitemOpen
  \bibfield  {author} {\bibinfo {author} {\bibfnamefont {R.}~\bibnamefont
  {Hajian}}\ and\ \bibinfo {author} {\bibfnamefont {S.}~\bibnamefont {Hardt}},\
  }\bibfield  {title} {\bibinfo {title} {Formation and lateral migration of
  nanodroplets via solvent shifting in a microfluidic device},\ }\href@noop {}
  {\bibfield  {journal} {\bibinfo  {journal} {Microfluid Nanofluid}\ }\textbf
  {\bibinfo {volume} {19}},\ \bibinfo {pages} {1281} (\bibinfo {year}
  {2015})}\BibitemShut {NoStop}%
\bibitem [{\citenamefont {Lu}\ \emph {et~al.}(2015)\citenamefont {Lu},
  \citenamefont {Xu}, \citenamefont {Zeng},\ and\ \citenamefont
  {Zhang}}]{Lu2015}%
  \BibitemOpen
  \bibfield  {author} {\bibinfo {author} {\bibfnamefont {Z.}~\bibnamefont
  {Lu}}, \bibinfo {author} {\bibfnamefont {H.}~\bibnamefont {Xu}}, \bibinfo
  {author} {\bibfnamefont {H.}~\bibnamefont {Zeng}},\ and\ \bibinfo {author}
  {\bibfnamefont {X.}~\bibnamefont {Zhang}},\ }\bibfield  {title} {\bibinfo
  {title} {Solvent effects on the formation of surface nanodroplets by solvent
  exchange},\ }\href@noop {} {\bibfield  {journal} {\bibinfo  {journal}
  {Langmuir}\ }\textbf {\bibinfo {volume} {31}},\ \bibinfo {pages}
  {12120–12125} (\bibinfo {year} {2015})}\BibitemShut {NoStop}%
\bibitem [{\citenamefont {Lu}\ \emph {et~al.}(2016)\citenamefont {Lu},
  \citenamefont {Peng},\ and\ \citenamefont {Zhang}}]{Lu2016}%
  \BibitemOpen
  \bibfield  {author} {\bibinfo {author} {\bibfnamefont {Z.}~\bibnamefont
  {Lu}}, \bibinfo {author} {\bibfnamefont {S.}~\bibnamefont {Peng}},\ and\
  \bibinfo {author} {\bibfnamefont {X.}~\bibnamefont {Zhang}},\ }\bibfield
  {title} {\bibinfo {title} {Influence of solution composition on the formation
  of surface nanodroplets by solvent exchange},\ }\href@noop {} {\bibfield
  {journal} {\bibinfo  {journal} {Langmuir}\ }\textbf {\bibinfo {volume}
  {32}},\ \bibinfo {pages} {1700–1706} (\bibinfo {year} {2016})}\BibitemShut
  {NoStop}%
\bibitem [{\citenamefont {Lu}\ \emph {et~al.}(2017)\citenamefont {Lu},
  \citenamefont {Schaarsberg}, \citenamefont {Zhu}, \citenamefont {Yeo},
  \citenamefont {Lohse},\ and\ \citenamefont {Zhang}}]{Lu2017}%
  \BibitemOpen
  \bibfield  {author} {\bibinfo {author} {\bibfnamefont {Z.}~\bibnamefont
  {Lu}}, \bibinfo {author} {\bibfnamefont {M.~H.~K.}\ \bibnamefont
  {Schaarsberg}}, \bibinfo {author} {\bibfnamefont {X.}~\bibnamefont {Zhu}},
  \bibinfo {author} {\bibfnamefont {L.~Y.}\ \bibnamefont {Yeo}}, \bibinfo
  {author} {\bibfnamefont {D.}~\bibnamefont {Lohse}},\ and\ \bibinfo {author}
  {\bibfnamefont {X.}~\bibnamefont {Zhang}},\ }\bibfield  {title} {\bibinfo
  {title} {Universal nanodroplet branches from confining the ouzo effect},\
  }\href@noop {} {\bibfield  {journal} {\bibinfo  {journal} {PNAS}\ }\textbf
  {\bibinfo {volume} {114}},\ \bibinfo {pages} {10332} (\bibinfo {year}
  {2017})}\BibitemShut {NoStop}%
\bibitem [{\citenamefont {Li}\ \emph {et~al.}(2018)\citenamefont {Li},
  \citenamefont {Bao}, \citenamefont {Yu},\ and\ \citenamefont
  {Zhang}}]{Li2018}%
  \BibitemOpen
  \bibfield  {author} {\bibinfo {author} {\bibfnamefont {M.}~\bibnamefont
  {Li}}, \bibinfo {author} {\bibfnamefont {L.}~\bibnamefont {Bao}}, \bibinfo
  {author} {\bibfnamefont {H.}~\bibnamefont {Yu}},\ and\ \bibinfo {author}
  {\bibfnamefont {X.}~\bibnamefont {Zhang}},\ }\bibfield  {title} {\bibinfo
  {title} {Formation of multicomponent surface nanodroplets by solvent
  exchange},\ }\href@noop {} {\bibfield  {journal} {\bibinfo  {journal} {J.
  Phys. Chem. C}\ }\textbf {\bibinfo {volume} {122}},\ \bibinfo {pages}
  {8647–8654} (\bibinfo {year} {2018})}\BibitemShut {NoStop}%
\bibitem [{\citenamefont {Dyett}\ \emph {et~al.}(2018)\citenamefont {Dyett},
  \citenamefont {Kiyama}, \citenamefont {Rump}, \citenamefont
  {Yoshiyuki~Tagawa},\ and\ \citenamefont {Zhang}}]{Dyett2018}%
  \BibitemOpen
  \bibfield  {author} {\bibinfo {author} {\bibfnamefont {B.}~\bibnamefont
  {Dyett}}, \bibinfo {author} {\bibfnamefont {A.}~\bibnamefont {Kiyama}},
  \bibinfo {author} {\bibfnamefont {M.}~\bibnamefont {Rump}}, \bibinfo {author}
  {\bibfnamefont {D.~L.}\ \bibnamefont {Yoshiyuki~Tagawa}},\ and\ \bibinfo
  {author} {\bibfnamefont {X.}~\bibnamefont {Zhang}},\ }\bibfield  {title}
  {\bibinfo {title} {Growth dynamics of surface nanodroplets during solvent
  exchange at varying flow rates},\ }\href@noop {} {\bibfield  {journal}
  {\bibinfo  {journal} {Soft Matter}\ }\textbf {\bibinfo {volume} {14}},\
  \bibinfo {pages} {5197} (\bibinfo {year} {2018})}\BibitemShut {NoStop}%
\bibitem [{\citenamefont {Zeng}\ \emph {et~al.}(2019)\citenamefont {Zeng},
  \citenamefont {Wang}, \citenamefont {Zhang},\ and\ \citenamefont
  {Lohse}}]{Zeng2019}%
  \BibitemOpen
  \bibfield  {author} {\bibinfo {author} {\bibfnamefont {B.}~\bibnamefont
  {Zeng}}, \bibinfo {author} {\bibfnamefont {Y.}~\bibnamefont {Wang}}, \bibinfo
  {author} {\bibfnamefont {X.}~\bibnamefont {Zhang}},\ and\ \bibinfo {author}
  {\bibfnamefont {D.}~\bibnamefont {Lohse}},\ }\bibfield  {title} {\bibinfo
  {title} {Solvent exchange in a {{Hele-Shaw}} cell: Universality of surface
  nanodroplet nucleation},\ }\href@noop {} {\bibfield  {journal} {\bibinfo
  {journal} {J. Phys. Chem. C}\ }\textbf {\bibinfo {volume} {123}},\ \bibinfo
  {pages} {5571–5577} (\bibinfo {year} {2019})}\BibitemShut {NoStop}%
\bibitem [{\citenamefont {Li}\ \emph {et~al.}(2021)\citenamefont {Li},
  \citenamefont {Chong}, \citenamefont {Bazyar}, \citenamefont {Lammertink},\
  and\ \citenamefont {Lohse}}]{Li2021}%
  \BibitemOpen
  \bibfield  {author} {\bibinfo {author} {\bibfnamefont {Y.}~\bibnamefont
  {Li}}, \bibinfo {author} {\bibfnamefont {K.}~\bibnamefont {Chong}}, \bibinfo
  {author} {\bibfnamefont {H.}~\bibnamefont {Bazyar}}, \bibinfo {author}
  {\bibfnamefont {R.}~\bibnamefont {Lammertink}},\ and\ \bibinfo {author}
  {\bibfnamefont {D.}~\bibnamefont {Lohse}},\ }\bibfield  {title} {\bibinfo
  {title} {Universality in microdroplet nucleation during solvent exchange in
  {{Hele-Shaw}}-like channels},\ }\href@noop {} {\bibfield  {journal} {\bibinfo
   {journal} {J. Fluid Mech.}\ }\textbf {\bibinfo {volume} {912}},\ \bibinfo
  {pages} {A35} (\bibinfo {year} {2021})}\BibitemShut {NoStop}%
\bibitem [{\citenamefont {Vaux}\ \emph {et~al.}(2019)\citenamefont {Vaux},
  \citenamefont {Mehaddi}, \citenamefont {Vauquelin},\ and\ \citenamefont
  {Candelier}}]{Vaux2019}%
  \BibitemOpen
  \bibfield  {author} {\bibinfo {author} {\bibfnamefont {S.}~\bibnamefont
  {Vaux}}, \bibinfo {author} {\bibfnamefont {R.}~\bibnamefont {Mehaddi}},
  \bibinfo {author} {\bibfnamefont {O.}~\bibnamefont {Vauquelin}},\ and\
  \bibinfo {author} {\bibfnamefont {F.}~\bibnamefont {Candelier}},\ }\bibfield
  {title} {\bibinfo {title} {Upward versus downward non-{B}oussinesq turbulent
  fountains},\ }\href@noop {} {\bibfield  {journal} {\bibinfo  {journal} {J.
  Fluid Mech.}\ }\textbf {\bibinfo {volume} {867}},\ \bibinfo {pages} {374}
  (\bibinfo {year} {2019})}\BibitemShut {NoStop}%
\bibitem [{\citenamefont {Hunt}\ and\ \citenamefont
  {Debugne}(2016)}]{Hunt2016}%
  \BibitemOpen
  \bibfield  {author} {\bibinfo {author} {\bibfnamefont {G.}~\bibnamefont
  {Hunt}}\ and\ \bibinfo {author} {\bibfnamefont {A.}~\bibnamefont {Debugne}},\
  }\bibfield  {title} {\bibinfo {title} {Forced fountains},\ }\href@noop {}
  {\bibfield  {journal} {\bibinfo  {journal} {J. Fluid Mech.}\ }\textbf
  {\bibinfo {volume} {802}},\ \bibinfo {pages} {437} (\bibinfo {year}
  {2016})}\BibitemShut {NoStop}%
\bibitem [{\citenamefont {Mehaddi}\ \emph {et~al.}(2015)\citenamefont
  {Mehaddi}, \citenamefont {Vauquelin},\ and\ \citenamefont
  {Candelier}}]{Mehaddi2015}%
  \BibitemOpen
  \bibfield  {author} {\bibinfo {author} {\bibfnamefont {R.}~\bibnamefont
  {Mehaddi}}, \bibinfo {author} {\bibfnamefont {O.}~\bibnamefont {Vauquelin}},\
  and\ \bibinfo {author} {\bibfnamefont {F.}~\bibnamefont {Candelier}},\
  }\bibfield  {title} {\bibinfo {title} {Experimental non-{B}oussinesq
  fountains},\ }\href@noop {} {\bibfield  {journal} {\bibinfo  {journal} {J.
  Fluid Mech.}\ }\textbf {\bibinfo {volume} {784}},\ \bibinfo {pages} {R6}
  (\bibinfo {year} {2015})}\BibitemShut {NoStop}%
\end{thebibliography}%

\end{document}